\documentclass[twocolumn]{aastex631}
\usepackage{graphicx,amsmath,amsthm}

\shorttitle{Weakened Magnetic Braking}
\shortauthors{Metcalfe et al.}

\newcommand\HD{HD\,76151}
\newcommand\Sco{18~Sco}
\newcommand\CenA{$\alpha$~Cen~A}
\newcommand\CygAB{16~Cyg~A~\&~B}

\journalinfo{The Astrophysical Journal Letters}

\begin{document}

\title{\large The Origin of Weakened Magnetic Braking in Old Solar Analogs}

\author[0000-0003-4034-0416]{Travis S.~Metcalfe}
\affiliation{White Dwarf Research Corporation, 9020 Brumm Trail, Golden, CO 80403, USA}

\author[0000-0002-3020-9409]{Adam J.~Finley}
\affiliation{\mbox{Department of Astrophysics-AIM, University of Paris-Saclay and University of Paris, CEA, CNRS, Gif-sur-Yvette Cedex F-91191, France}}

\author[0000-0003-3061-4591]{Oleg Kochukhov}
\affiliation{Department of Physics and Astronomy, Uppsala University, Box 516, SE-75120 Uppsala, Sweden}

\author[0000-0001-5986-3423]{Victor~See}
\affiliation{\mbox{European Space Agency (ESA), European Space Research and Technology Centre (ESTEC), Keplerlaan 1, 2201 AZ Noordwijk, the Netherlands}}

\author[0000-0002-1242-5124]{Thomas R.~Ayres}
\affiliation{Center for Astrophysics and Space Astronomy, 389 UCB, University of Colorado, Boulder, CO 80309, USA}

\author[0000-0002-3481-9052]{Keivan G.~Stassun}
\affiliation{Vanderbilt University, Department of Physics \& Astronomy, 6301 Stevenson Center Lane, Nashville, TN 37235, USA}

\author[0000-0002-4284-8638]{Jennifer L.~van~Saders}
\affiliation{Institute for Astronomy, University of Hawai`i, 2680 Woodlawn Drive, Honolulu, HI 96822, USA}

\author[0000-0002-2361-5812]{Catherine A.~Clark}
\affiliation{Northern Arizona University, 527 South Beaver Street, Flagstaff, AZ 86011, USA}
\affiliation{Lowell Observatory, 1400 West Mars Hill Road, Flagstaff, AZ 86001, USA}

\author[0000-0003-4556-1277]{Diego Godoy-Rivera}
\affiliation{Department of Astronomy, The Ohio State University, 140 West 18th Avenue, Columbus, OH 43210, USA}
\affiliation{Instituto de Astrof\'{\i}sica de Canarias, E-38205 La Laguna, Tenerife, Spain}
\affiliation{Universidad de La Laguna, Departamento de Astrofísica, E-38206 La Laguna, Tenerife, Spain}

\author[0000-0002-0551-046X]{Ilya V.~Ilyin}
\affiliation{Leibniz-Institut f\"ur Astrophysik Potsdam (AIP), An der Sternwarte 16, D-14482 Potsdam, Germany}

\author[0000-0002-7549-7766]{Marc H.~Pinsonneault}
\affiliation{Department of Astronomy, The Ohio State University, 140 West 18th Avenue, Columbus, OH 43210, USA}

\author[0000-0002-6192-6494]{Klaus G.~Strassmeier}
\affiliation{Leibniz-Institut f\"ur Astrophysik Potsdam (AIP), An der Sternwarte 16, D-14482 Potsdam, Germany}

\author[0000-0001-7624-9222]{Pascal Petit}
\affiliation{Universit\'e de Toulouse, CNRS, CNES, 14 avenue Edouard Belin, 31400, Toulouse, France}

\begin{abstract}

The rotation rates of main-sequence stars slow over time as they gradually lose angular 
momentum to their magnetized stellar winds. The rate of angular momentum loss depends on 
the strength and morphology of the magnetic field, the mass-loss rate, and the stellar 
rotation period, mass, and radius. Previous observations suggested a shift in magnetic 
morphology between two F-type stars with similar rotation rates but very different ages 
(88~Leo and $\rho$~CrB). In this Letter, we identify a comparable transition in an 
evolutionary sequence of solar analogs with ages between 2--7~Gyr. We present new 
spectropolarimetry of \Sco\ and \CygAB\ from the Large Binocular Telescope, and we 
reanalyze previously published Zeeman Doppler images of \HD\ and \Sco, providing 
additional constraints on the nature and timing of this transition. We combine archival 
X-ray observations with updated distances from Gaia to estimate mass-loss rates, and we 
adopt precise stellar properties from asteroseismology and other sources. We then 
calculate the wind braking torque for each star in the evolutionary sequence, 
demonstrating that the rate of angular momentum loss drops by more than an order of 
magnitude between the ages of \HD\ and \Sco\ (2.6--3.7~Gyr) and continues to decrease 
modestly to the age of \CygAB\ (7~Gyr). We suggest that this magnetic transition may 
represent a disruption of the global dynamo arising from weaker differential rotation, 
and we outline plans to probe this phenomenon in additional stars spanning a wide range 
of spectral types.

\end{abstract}


\section{Introduction}\label{sec1}

Fifty years after \cite{Skumanich1972} used the Sun and a few young star clusters to 
infer that rotation slows with the square-root of age, it is clear that rotational 
evolution is anything but smooth in solar-type stars. Detailed characterization of some 
of the same clusters considered by Skumanich recently identified a mass-dependent pause 
in the spin-down of stars that can last from 100~Myr to a few Gyr \citep{Curtis2019, 
Curtis2020}. This pause may signal a redistribution of angular momentum from the 
radiative interior to the convective envelope \citep{Spada2020}, temporarily compensating 
for the gradual loss of angular momentum to magnetized stellar winds. The extrapolation 
of the Skumanich spin-down relation beyond the solar age also appears to be unreliable, 
with old field stars observed by the Kepler space telescope rotating substantially faster 
than expected \citep{Angus2015}. This behavior may reflect a weakening of magnetic 
braking beyond a critical Rossby number ($\textrm{Ro} \equiv P_{\rm rot}/\tau_c$), when 
the ratio of the rotation period to the convective overturn timescale becomes comparable 
to the solar value \citep{vanSaders2016, vanSaders2019}. This conclusion was reinforced
by stars with ages and rotation rates both determined from asteroseismology 
\citep{Hall2021}, and the resulting accumulation of stars with a broad range of ages at 
similar rotation periods was identified in stellar samples with precise effective 
temperatures \citep{David2022}.

The evidence for a reduced torque is powerful, but to understand the cause we need to 
tie it to a physical mechanism. Weakened magnetic braking in middle-aged stars may 
result from either a shift in magnetic morphology from larger to smaller spatial scales 
\citep{Reville2015, Garraffo2016}, an abrupt change in the mass-loss rate 
\citep{Vidotto2018}, or both. The global dipole can sustain more open magnetic flux 
where stellar wind can escape---giving it the longest effective lever arm---so a 
disruption of the largest-scale organization of the magnetic field can dramatically 
reduce the efficiency of angular momentum loss. However, very few constraints are 
available for old inactive stars from existing spectropolarimetric data sets 
\citep[e.g.,][]{See2019}. Additional observations of specific evolutionary sequences of 
stars can help elucidate the origin of weakened magnetic braking. \cite{Finley2018} 
used a set of magnetohydrodynamic wind simulations to develop a parametrization of the 
wind braking torque based on the strength and morphology of the magnetic field, the 
mass-loss rate, and the stellar rotation period, mass, and radius. By comparing stars 
along an evolutionary sequence, this parametrization allows us to evaluate the relative 
importance of various contributions to the total change in angular momentum loss. 
\cite{Metcalfe2019, Metcalfe2021} used this approach to identify a shift in magnetic 
morphology between two F-type stars with similar rotation rates but very different ages 
(88~Leo and $\rho$~CrB).

In this Letter, we identify a comparable transition in an evolutionary sequence of solar 
analogs with ages between 2--7~Gyr. In Section~\ref{sec2}, we describe the new and 
archival observations that allow us to constrain the input parameters of the wind braking 
prescription. In Section~\ref{sec3}, we calculate the rate of angular momentum loss for 
each of the stars in the evolutionary sequence, and we determine the most important 
changes that influence the wind braking torque at different ages. Finally, in 
Section~\ref{sec4} we speculate on the physical mechanisms that might be responsible for 
the inferred magnetic transition, and we outline future tests that can further 
characterize the nature and timing of this phenomenon.

 \begin{figure*}
 \centerline{\includegraphics[width=3.5in,trim=5 30 25 30,clip]{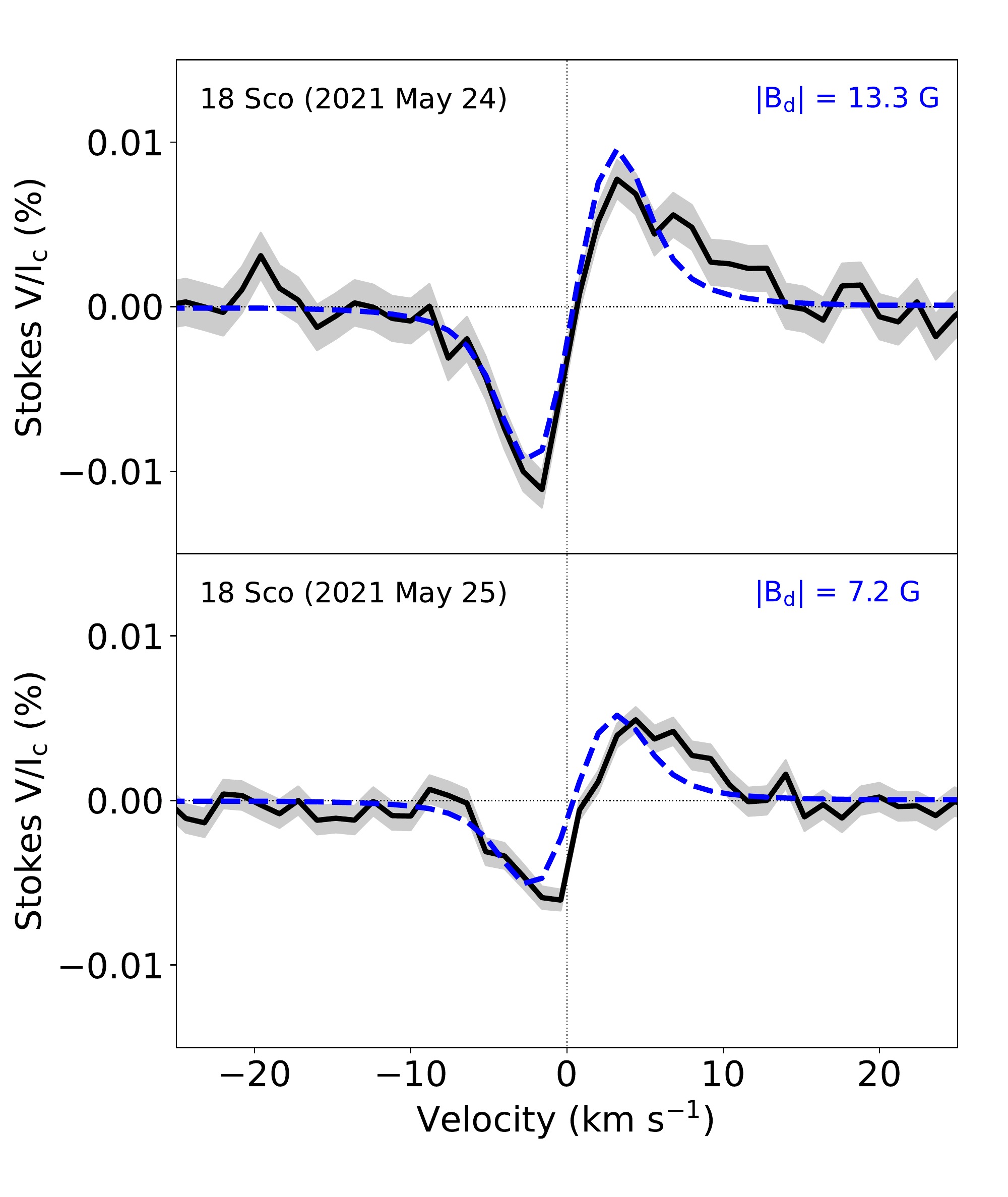}\includegraphics[width=3.5in,trim=5 30 25 30,clip]{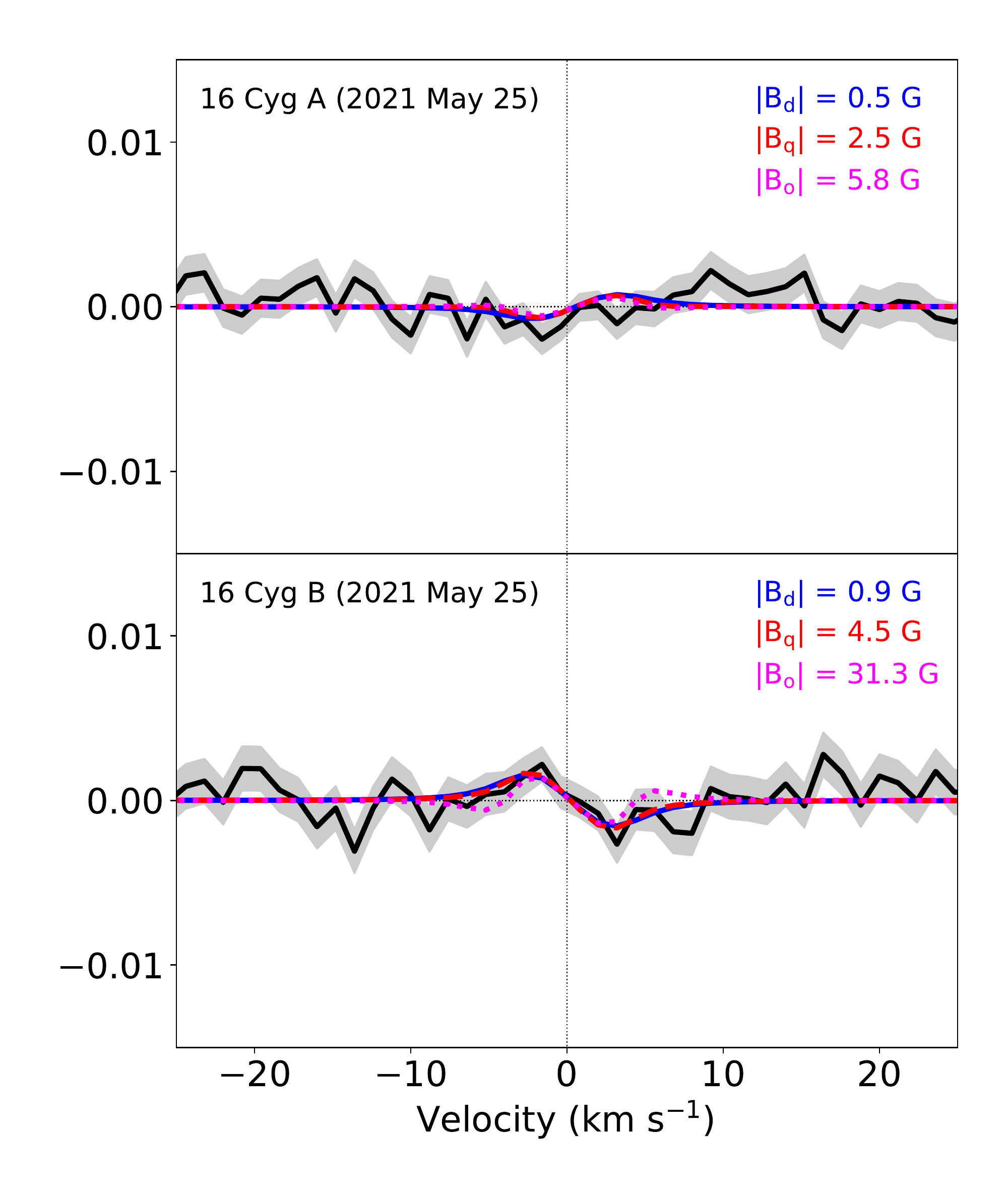}}
 \caption{Stokes~$V$ circular polarization profiles for \Sco\ (left) and \CygAB\ (right) 
from LBT observations in 2021 May. Mean profiles are shown as black lines, while 
uncertainties are indicated with gray shaded areas. Colored lines show axisymmetric model 
profiles assuming dipole (blue), quadrupole (red), or octupole (magenta) geometry with 
fixed inclination.\\ \label{fig1}}
 \end{figure*}

\section{Observations}\label{sec2}

\subsection{Spectropolarimetry}\label{sec2.1}

We observed \Sco\ and \CygAB\ in 2021 May at $R$\,=\,130,000 with the Potsdam 
Echelle Polarimetric and Spectroscopic Instrument \citep[PEPSI;][]{Strassmeier2015} on 
the 2$\times$8.4~m Large Binocular Telescope (LBT), using the same instrumental setup 
and data reduction methods described in \cite{Metcalfe2019}. We applied the 
least-squares deconvolution technique \citep[LSD;][]{Kochukhov2010} to these 
observations in order to derive precise mean intensity and polarization profiles. The 
line data required for the LSD analysis were obtained from the VALD database 
\citep{Ryabchikova2015}, adopting stellar atmospheric parameters from \cite{Brewer2016}. 
By combining information from between \mbox{1150--1300} metal lines deeper than 10\% of 
the continuum, we obtained mean Stokes~$V$ profiles with an uncertainty of 
$\sim$10$^{-5}$ (see Figure~\ref{fig1}). The data yielded a definite detection of the 
polarization signature for \Sco\ on both observing nights, with a mean longitudinal 
magnetic field $\langle B_{\rm z}\rangle$=$-2.15\pm0.23$~G and $-1.06\pm0.14$~G on 2021 
May 24 and 25, respectively. These Stokes~$V$ profiles are similar to those 
obtained by \cite{Petit2008}, but with a much higher signal-to-noise ratio. No 
significant polarization signatures were detected in the LSD profiles for \CygAB, which 
yielded $\langle B_{\rm z}\rangle$=$-0.27\pm0.13$~G and $+0.16\pm0.15$~G, respectively. 
Following the line profile modeling methodology described in \cite{Metcalfe2019}, we 
estimated upper limits on the strengths of axisymmetric dipole, quadrupole, and octupole 
magnetic fields that are compatible with the observed polarization profiles for \CygAB\ 
(see Table~\ref{tab1}).

To complement these new observations, we reanalyzed the Zeeman Doppler imaging (ZDI) 
maps for \HD\ and \Sco\ published in \cite{Petit2008}, which were obtained near the mean 
activity level and near cycle maximum, respectively. These ZDI maps were obtained 
with an $R$\,=\,65,000 spectropolarimeter attached to a 2~m telescope. The wind braking 
prescription of \cite{Finley2018} was constructed using simulations that only contained 
axisymmetric magnetic fields. However, the observed ZDI maps contain both axisymmetric 
and non-axisymmetric components. We therefore need a method of calculating the 
equivalent axisymmetric dipole, quadrupole, and octupole fields to use in the wind 
braking prescription for these stars. This can be done by considering the magnetic flux 
of each component of the ZDI map, since it is the radial dependence of the magnetic flux 
that is important for determining angular momentum loss. Specifically, we calculated the 
magnetic flux associated with just the dipole field in each ZDI map, accounting for both 
axisymmetric and non-axisymmetric components. We then determined the polar field 
strength ($B_{\rm d}$) required for an axisymmetric dipole field to match the total 
magnetic flux associated with all dipole components in the ZDI map. We repeated this 
procedure for the analysis of the quadrupole ($B_{\rm q}$) and octupole ($B_{\rm o}$) 
components of the ZDI maps, and the resulting polar field strengths were all used 
for the wind braking torque calculations. The results are shown in Table~\ref{tab1}.

\begin{deluxetable*}{lcccc}
  \setlength{\tabcolsep}{18pt}
  \tablecaption{Stellar Properties of the Evolutionary Sequence\label{tab1}}
  \tablehead{\colhead{}            & \colhead{HD\,76151}    & \colhead{18~Sco}       & \colhead{16~Cyg~A}   & \colhead{16~Cyg~B}   }
  \startdata
  $T_{\rm eff}$ (K)                & $5790 \pm 44$          & $5785 \pm 25$          & $5778 \pm 25$        & $5747 \pm 25$        \\
  $[$M/H$]$ (dex)                  & $+0.07 \pm 0.03$       & $+0.04 \pm 0.01$       & $+0.09 \pm 0.01$     & $+0.06 \pm 0.01$     \\
  $\log g$ (dex)                   & $4.55 \pm 0.06$        & $4.41 \pm 0.03$        & $4.28 \pm 0.03$      & $4.37 \pm 0.03$      \\
  $B-V$ (mag)                      & $0.661$                & $0.652$                & $0.643$              & $0.661$              \\
  $\log R'_{\rm HK}$ (dex)         & $-4.698 \pm 0.014$     & $-4.857 \pm 0.020$     & $-5.074 \pm 0.014$   & $-5.041 \pm 0.014$   \\
  $P_{\rm rot}$ (days)             & $20.5 \pm 0.3$         & $22.7 \pm 0.5$         & $20.5^{+2.0}_{-1.1}$ & $21.2^{+1.8}_{-1.5}$ \\
  Inclination ($^\circ$)           & $30 \pm 15$            & $70^{+20}_{-25}$       & $45^{+6}_{-3}$       & $34 ^{+3}_{-2}$      \\
  $|B_{\rm d}|$ (G)                & $5.13$                 & $1.34$                 & $< 0.5$              & $< 0.9$              \\
  $|B_{\rm q}|$ (G)                & $2.88$                 & $2.01$                 & $< 2.5$              & $< 4.5$              \\
  $|B_{\rm o}|$ (G)                & $1.34$                 & $0.86$                 & $< 5.8$              & $< 31.3$             \\
  $L_X$ ($10^{27}$~erg~s$^{-1}$)   & $29 \pm 2$             & $1.7 \pm 0.7$          & $2.7 \pm 0.5$        & $1.2 \pm 0.3$        \\
  Mass-loss rate ($\dot{M}_\odot$) & $8.3 \pm 0.7$          & $0.87 \pm 0.32$        & $0.92 \pm 0.16$      & $0.57 \pm 0.13$      \\
  Mass ($M_\odot$)                 & $1.05 \pm 0.06$        & $1.02 \pm 0.03$        & $1.072 \pm 0.013$    & $1.038 \pm 0.047$    \\
  Radius ($R_\odot$)               & $0.964 \pm 0.018$      & $1.010 \pm 0.009$      & $1.223 \pm 0.005$    & $1.113 \pm 0.016$    \\
  Luminosity ($L_\odot$)           & $0.938 \pm 0.022$      & $1.067 \pm 0.032$      & $1.52 \pm 0.05$      & $1.21 \pm 0.11$      \\
  Age (Gyr)                        & $2.6 \pm 0.4$          & $3.66^{+0.44}_{-0.50}$ & $7.36 \pm 0.31$      & $7.05 \pm 0.63$      \\
  Sources                          & 1, 3, 4, 7, 8, 9       & 2, 3, 4, 7, 10, 11     & 2, 3, 5, 6, 7, 12    & 2, 3, 5, 6, 7, 12    \\
  \hline
  Torque ($10^{30}$~erg)           & $4.77$                 & $0.378$                & $<0.316$--0.494      & $<0.302$--0.375      \\
  \enddata
  \tablerefs{(1)~\cite{Valenti2005}; (2)~\cite{Brewer2016}; (3)~\cite{Radick2018}; (4)~\cite{Petit2008}; (5)~\cite{Hall2021}; (6)~Section\,\ref{sec2.1}; (7)~Section\,\ref{sec2.2}; (8)~Section\,\ref{sec2.3}; (9)~\cite{Barnes2007}; (10)~\cite{Bazot2011}; (11)~\cite{Li2012}; (12)~\cite{Creevey2017}}
\vspace*{-18pt}
\end{deluxetable*}
\vspace*{-24pt}

\subsection{X-ray Data}\label{sec2.2}

We determined the X-ray luminosity ($L_X$) for each of our targets following the 
procedures described in \cite{Ayres2022}, using archival observations from several 
missions. The ROSAT All-Sky Bright Source Catalog \citep{Boller2016} included 
observations of \HD\ from 1990--1991 (slightly below the mean activity level), while 
Chandra observed \Sco\ with the \mbox{ACIS-S} instrument in 2011 (ObsID 12393, near 
magnetic minimum) and \CygAB\ using the \mbox{HRC-I} instrument in 2020 (ObsID 21167 \& 
23188, constant activity). All four targets were identified in a catalog of serendipitous 
point sources observed by \mbox{XMM-Newton} \citep{Traulsen2020}, providing an 
independent estimate of the X-ray luminosity when converted to the standard 0.1--2.4~keV 
energy band. The modeling approach entails convolving the empirical emission measure 
distributions (EMD) from \cite{Wood2018} with temperature-dependent energy conversion 
factors (ECF) for each particular instrument. The EMD models correspond to a range of 
X-ray surface fluxes, while each model-derived ECF implies an empirical surface flux, 
given the net count rate of the observation and the stellar parameters based on Gaia EDR3 
\citep{Gaia2021}. When the predicted surface flux agrees with the input model value, this 
indicates the self-consistent choice for the ECF. After modeling each individual 
observation, we adopted the consensus estimates shown in Table~\ref{tab1}. The 
conservative errors are intended to reflect possible temporal variations due to activity 
cycles as well as inter-instrumental calibration inconsistencies, including uncertainties 
in energy band conversion factors. We estimated mass-loss rates using the scaling 
relation \mbox{$\dot{M}\propto F_X^{0.77\pm0.04}$} \citep{Wood2021}, calculating surface 
fluxes from the X-ray luminosities and stellar radii shown in Table~\ref{tab1}.

\subsection{Stellar Properties}\label{sec2.3}

In addition to the magnetic field strength and morphology from spectropolarimetry and the 
mass-loss rate from X-ray fluxes, the other parameters that are required to calculate the 
wind braking torque are the stellar rotation period, mass, and radius. For \HD\ and \Sco\ 
we adopted the rotation periods and inclinations from \cite{Petit2008}, while for \CygAB\ 
we adopted the asteroseismic determinations of both parameters from \cite{Hall2021}. 
Aside from \HD, stellar masses and radii for our targets are available from 
asteroseismology. For \Sco\ we adopted the values from \cite{Bazot2011}, while for 
\CygAB\ we adopted those from \cite{Creevey2017}. For \HD\ we determined the stellar 
radius from an analysis of the spectral energy distribution, following 
\cite{Metcalfe2021} with procedures adapted from \cite{Stassun:2017, Stassun:2018}, and 
we estimated the stellar mass from the empirical eclipsing-binary-based relations of 
\cite{Torres2010}. Stellar ages are from gyrochronology for \HD\ \citep{Barnes2007}, and 
asteroseismology for \Sco\ \citep{Li2012} and \CygAB\ \citep{Creevey2017}. 
Table~\ref{tab1} includes a complete listing of the adopted and computed stellar 
properties.

\section{Wind Braking Torque}\label{sec3}

With all of the observational inputs determined, we can now estimate the wind braking 
torque for each of the stars in the evolutionary sequence using the prescription of 
\cite{Finley2018}\footnote{\url{https://github.com/travismetcalfe/FinleyMatt2018}}. For \HD\ 
and \Sco\ we have the equivalent polar field strengths $(B_{\rm d}, B_{\rm q}, B_{\rm 
o})$ derived in Section~\ref{sec2.1} from the ZDI maps. For \CygAB\ we only have upper 
limits, each of which assumes that all of the field is in a single component, so we 
calculate the range of torques resulting from the three upper limits on polar field 
strength. Combining the information from spectropolarimetry with the other stellar 
properties $(\dot M, P_{\rm rot}, M, R)$, we find that the wind braking torque for \Sco\ 
(3.7~Gyr) and \CygAB\ (7~Gyr) are all more than an order of magnitude smaller than for 
\HD\ (2.6~Gyr). The calculated wind braking torques are listed in Table~\ref{tab1}. 
The large change in wind braking torque between \HD\ and \Sco\ is comparable to the 
inferred change between the F-type stars 88~Leo (2.4~Gyr) and $\rho$~CrB (9.8~Gyr) by 
\cite{Metcalfe2021} at roughly the same Rossby number. Identical methods applied to  
these stars yields a torque of $4.13 \times 10^{30}$~erg for 88~Leo, and a range of 
upper limits $<0.296$--0.337 $\times\ 10^{30}$~erg for $\rho$~CrB.

Evolutionary changes in magnetic field strength, mass-loss rate, rotation period, and 
radius are all expected on Gyr timescales, so it's important to evaluate the influence of 
each input parameter on the total change in wind braking torque. Following 
\cite{Metcalfe2021}, we assess the various contributions by modifying one parameter at a 
time between the adopted values for two stars along the evolutionary sequence, ranking 
their influence by the associated decrease in the absolute wind braking torque. The 
magnetic field strength is indicated by the absolute values of $(B_{\rm d}, B_{\rm q}, 
B_{\rm o})$, while the field morphology is reflected in their relative values. These two 
contributions are not easily separable without additional assumptions, so we calculate 
their combined influence by changing $(B_{\rm d}, B_{\rm q}, B_{\rm o})$ 
simultaneously. Comparing the fiducial models for \HD\ and \Sco, we find that the change 
in wind braking torque (\mbox{2.6--3.7~Gyr}) is dominated by the evolution of the 
mass-loss rate ($-$69\%) and magnetic field strength and morphology ($-$65\%), with 
additional contributions from differences in the rotation period ($-$10\%) and stellar 
radius ($-$4\%). The small difference in stellar mass between the two stars slightly 
offsets (+0.4\%) the overall decrease in wind braking torque.

To isolate the evolutionary changes that occur between the ages of \Sco\ and \CygAB, we 
repeated the procedure above using the fiducial models for these stars. For simplicity, 
we adopt the upper limits on $B_{\rm d}$ for \CygAB\ with the remaining field components 
set to zero, yielding the strongest constraints on the wind braking torque. Once again we 
rank the influence of each parameter by the decrease in absolute wind braking torque, 
grouping the associated changes in parentheses for \CygAB, respectively. We find that 
beyond the age of \Sco\ (\mbox{3.7--7~Gyr}) the wind braking torque continues to weaken 
modestly for \CygAB\ ($-$17\%, $-$20\%). Importantly, the decrease in wind braking torque 
becomes dominated by the evolution of magnetic field strength and morphology ($-$59\%, 
$-$31\%), with smaller contributions from differences in the mass-loss rate (+3\%, 
$-$20\%) and stellar mass ($-$1\%, $-$0.4\%). These reductions are substantially offset 
by evolutionary changes in stellar radius (+83\%, +36\%) and by small differences in 
rotation period (+11\%, +7\%).

We can forward model the rotation periods of the evolutionary sequence to provide 
supporting evidence for weakened magnetic braking near the age of \Sco, following the 
same approach described in \cite{Metcalfe2021}. Using $T_{\rm eff}$, [M/H], and radius as 
constraints, and adopting priors on mass, age, and bulk metallicity, the standard 
spin-down model predicts rotation periods of $18\pm2$, $23\pm2$, $41\pm2$, and 
$38^{+2}_{-3}$~days for \HD, \Sco, \CygAB, respectively. By contrast, the weakened 
magnetic braking model predicts rotation periods of $18\pm2$, $22\pm1$, $26\pm1$, and 
$25\pm1$~days, much closer to the observed periods for \CygAB.

\section{Discussion}\label{sec4}

By combining spectropolarimetric measurements for an evolutionary sequence of solar 
analogs with the wind braking prescription of \cite{Finley2018}, we have 
placed new constraints on the nature and timing of weakened magnetic braking in 
middle-aged stars. In particular, we have shown that the wind braking torque decreases by 
more than an order of magnitude between the ages of \HD\ and \Sco\ (2.6--3.7~Gyr), 
substantially exceeding theoretical expectations. Angular momentum losses during this 
epoch are dominated by evolutionary changes in both the mass-loss rate and magnetic field 
strength and morphology. The wind braking torque continues to decrease modestly between 
the ages of \Sco\ and \CygAB\ (3.7--7~Gyr) primarily from changes in the magnetic field 
strength and morphology, despite being substantially offset by evolutionary changes in 
the stellar radius. Unfortunately, constraints on the magnetic morphology at the 
intermediate age of \CenA\ (5.4~Gyr) are relatively weak due to an unfavorable stellar 
inclination \citep{Kochukhov2011}. Our results corroborate previous indications of a 
magnetic transition from F-type stars \citep{Metcalfe2019, Metcalfe2021}.

With growing observational evidence of a magnetic transition in middle-aged stars, we can 
now speculate on the physical mechanisms that might be responsible. Large-scale 
magnetic fields in stars are thought to be generated by a global dynamo process, in which 
differential rotation provides the shear to transform poloidal field into toroidal field 
\citep{Charbonneau2020}. \cite{Metcalfe2016} suggested a fundamental shift in the 
character of differential rotation for middle-aged stars, based on the observed 
concentration of field into the poloidal component for young solar analogs 
\citep{Petit2008}, as well as the theoretical transition to anti-solar differential 
rotation in global convection simulations at high Rossby number \citep{Gastine2014}. Some 
of the details of this suggestion were misguided by differences in how the Rossby number 
is calculated for three-dimensional simulations of convection compared to one-dimensional 
stellar models that adopt mixing-length theory \citep{Brun2017}, but qualitatively it 
remains a viable explanation. Subsequent observational evidence supporting this 
suggestion came from \cite{Benomar2018}, who attempted to measure latitudinal 
differential rotation from the asteroseismic mode splittings of Kepler targets and found 
two-thirds of the sample consistent with uniform rotation. The absence of 
significant differential rotation at high Rossby number in this sample is consistent with 
the notion that weaker differential rotation may be related to the observed onset of 
weakened magnetic braking. \cite{Bazot2019} applied a similar analysis to \CygAB\ and 
found weak differential rotation comparable to the Sun, but with the highest 
stellar latitudes largely unconstrained by observations. This limitation could be 
addressed with additional analysis of the octupole modes observed in these stars.

 \begin{figure}[t]
 \centerline{\includegraphics[width=\columnwidth,trim=10 0 20 15,clip]{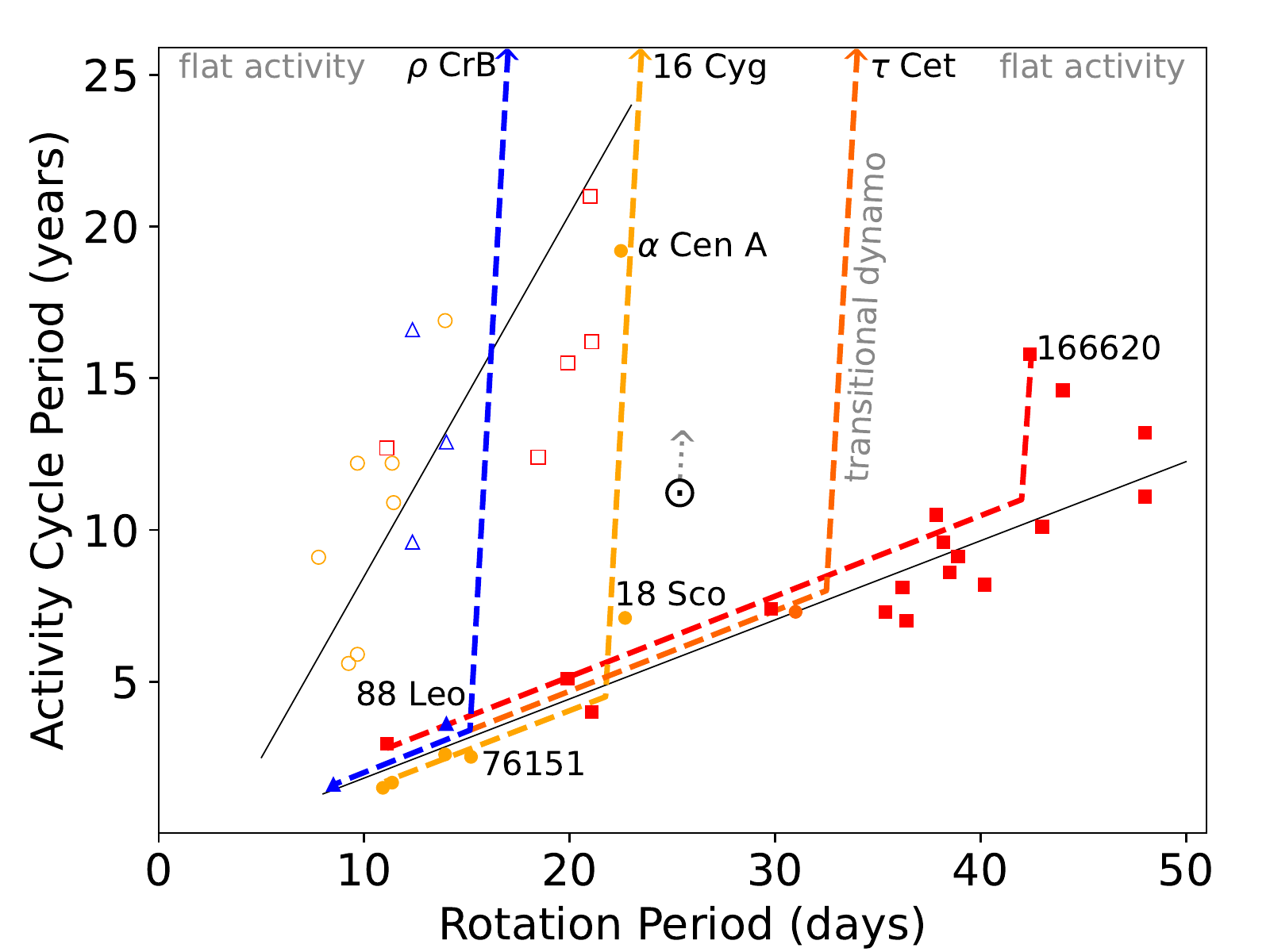}}
 \caption{Dependence of activity cycle period on rotation, showing two distinct sequences 
(solid lines). Points are colored by effective temperature, indicating F-type (blue 
triangles), G-type (yellow and orange circles), and K-type stars (red squares). Schematic 
evolutionary tracks are shown as dashed lines \citep[][and references 
therein]{Metcalfe2017}, leading to stars with constant activity that appear to have shut 
down their global dynamos (arrows along top).\label{fig2}}
 \end{figure}

If a diminishing large-scale field is responsible for weakened magnetic braking, then a 
coincident shift in the underlying dynamo might be found in observations of stellar 
activity cycles. \cite{Metcalfe2017} attempted to identify such a signature, and 
proposed an evolutionary thread connecting younger stars that show a clear relationship 
between rotation rate and activity cycle period (like \HD), to older stars that appear 
to be outliers from this relationship or exhibit constant activity on decadal 
timescales. The template for this evolutionary thread included the solar analogs \Sco\ 
at 3.7~Gyr, \CenA\ at 5.4~Gyr, and \CygAB\ near 7~Gyr (see Figure~\ref{fig2}). These 
stars all have comparable rotation periods, but the 7~yr activity cycle in \Sco\ 
\citep{Hall2007} will apparently become longer and weaker by the age of \CenA\ 
\citep{Ayres2014} before disappearing entirely by the age of \CygAB\ \citep{Radick2018}. 
Similar patterns were found in hotter stars with faster rotation and among cooler stars 
with slower rotation, at roughly the same critical Rossby number identified by 
\cite{vanSaders2016}. Within this context, the Maunder minimum in the Sun can be 
interpreted as intermittency in the solar cycle as it grows longer and weaker on stellar 
evolutionary timescales. Additional evidence for this interpretation comes from the 
recent observation of HD\,166620 apparently entering a grand minimum \citep{Baum2022}, 
after showing clear cyclic behavior during the Mount Wilson survey \citep{Baliunas1995}. 
This star has a Rossby number similar to the solar value and appears slightly above the 
so-called inactive sequence of stellar activity cycles, an outlier comparable to the Sun 
\citep{BohmVitense2007}.

We can further characterize the nature and timing of the magnetic transition by combining 
spectropolarimetry of carefully selected targets with reliable stellar ages from 
asteroseismology. We have already obtained new LBT snapshots of the solar analogs 
HD\,126053 and $\lambda$~Ser, and future observations will target $\tau$~Cet and 
HD\,166620 along with some active comparison stars at similar rotation periods (CF~UMa 
and 40~Eri). Most of these stars have recently been observed at 20-second cadence with 
the Transiting Exoplanet Survey Satellite \citep[TESS;][]{Ricker2014}, allowing a search 
for solar-like oscillations to help constrain the stellar properties \citep[e.g., 
see][]{Metcalfe2021}. Results from the initial reconnaissance with LBT will help guide 
the longer-term spectropolarimetry that is required to reconstruct complete Zeeman 
Doppler images for these targets, reducing the ambiguities that are inherent in snapshot 
observations. Within a few years, we should be able to apply the methodology outlined 
above to new observations spanning a wide range of spectral types, which sample 
different convective overturn timescales.\\

\noindent 
Special thanks to Wayne Rosing and Kenneth Kulju for supporting this project, 
and to Sean Matt and Keith MacGregor for helpful suggestions. T.S.M.\ acknowledges 
support from the Vanderbilt Initiative in Data-intensive Astrophysics (VIDA). A.J.F.\ is 
supported by the ERC Synergy grant ``Whole Sun,'' No.~810218. O.K.\ acknowledges support 
by the Swedish Research Council (project 2019-03548). V.S.\ acknowledges support from the 
European Space Agency (ESA) as an ESA Research Fellow. The LBT is an international 
collaboration among institutions in the United States, Italy and Germany. LBT Corporation 
partners are: The University of Arizona on behalf of the Arizona Board of Regents; 
Istituto Nazionale di Astrofisica, Italy; LBT Beteiligungsgesellschaft, Germany, 
representing the Max-Planck Society, The Leibniz Institute for Astrophysics Potsdam, and 
Heidelberg University; The Ohio State University, and The Research Corporation, on behalf 
of The University of Notre Dame, University of Minnesota and University of Virginia.


\end{document}